\documentclass{article}

\usepackage{amsmath, amsthm, amsfonts}
\usepackage{authblk}
\usepackage{multirow}
\usepackage{graphicx}

\theoremstyle{definition}

\theoremstyle{remark}

\title{New statistical model for misreported data with application to current public health challenges}
\author[1,2*]{David Mori\~na}
\author[3]{Amanda Fern\'andez-Fontelo}
\author[4]{Alejandra Caba\~na}
\author[2,4]{Pedro Puig}
\affil[1]{Department of Econometrics, Statistics and Applied Economics, Riskcenter-IREA, Universitat de Barcelona, Barcelona, Spain}
\affil[2]{Centre de Recerca Matem\`atica, Cerdanyola del Vall\`es, Spain}
\affil[3]{Chair of Statistics, Humboldt-Universit\"at zu Berlin, Berlin, Germany}
\affil[4]{Barcelona Graduate School of Mathematics (BGSMath), Departament de Matem\`atiques, Universitat Aut\`onoma de Barcelona, Cerdanyola del Vall\`es, Spain}
\affil[*]{dmorina@ub.edu}

\begin{document}
\maketitle

\abstract{The main goal of this work is to present a new model able to deal with potentially 
misreported continuous time series. The proposed model is able to handle the autocorrelation
structure in continuous time series data, which might be partially or totally underreported or overreported. Its performance is illustrated through a comprehensive simulation study considering several autocorrelation structures and two real data applications on human papillomavirus incidence in Girona (Catalunya, Spain) and Covid-19 incidence in the Chinese region of Heilongjiang.}

\section{Introduction}\label{intro}
There has been a growing interest in the past years to deal with data that is only partially registered or underreported 
in the time series literature. This phenomenon is very common in many fields, and has been previously explored by different 
approaches in epidemiology, social and biomedical research among many other contexts \cite{Bernard2014,Arendt2013,Rosenman2006,Alfonso2015,Winkelmann1996}. The sources and underlying
mechanisms that cause the underreporting might differ depending on the particular data. Some authors consider a situation where the registry is updated with time and therefore the underreporting issue is mitigated \cite{Hohle2014}. That leads to temporary underreporting while this work is focused on permanent underreporting, where the registered data are never updated in order to become more accurate. From the methodological point of view, several alternatives have been explored, from Markov chain Monte-Carlo based methods \cite{Winkelmann1996} to recent discrete time series approaches \cite{Fernandez-Fontelo2016,FernandezFontelo2019}. Several attempts to estimate the degree of underreporting in different contexts have been done \cite{Gibbons2014}, although there is a lack of models incorporating continuous time series structures and handling underreporting. 

One of the fields where the interest in addressing the underreporting issues is higher is the epidemiology of infectious diseases. In the last few years, many approaches to deal with underreported data have been suggested with a growing level of sophistication from the usage of multiplication factors \cite{Stocks2018} to several Markov-based models \cite{Azmon2014,Magal2018} or even spatio-temporal modelling \cite{Stoner2019}. Even a new R \cite{RCoreTeam2019} package able to fitting endemic-epidemic models based on approximative maximum likelihood to underreported count data has been recently published \cite{JohannesBracher2019}. This work presents two examples where such phenomenon appears.

Human papillomavirus (HPV) is one of the most prevalent sexually transmitted infections. It is so common that nearly all sexually active people have it at some point in their lives, according to the information provided by the United States' Centers for Disease Control and Prevention (CDC) \cite{Dunne2014}. Generally, the infection disappears on its own without inducing any health problem, but in some cases it can produce an abnormal growth of cells on the surface of the cervix that could potentially lead to cervical cancer. HPV infection is also related to other cancers (vulva, vagina, penis, anus, $\dots$) and other diseases like genital warts (GW). The fact that most cases of HPV infection are asymptomatic causes that public health registries might be potentially underestimating its incidence. The underreporting phenomenon in HPV data from the discrete time series point of view has been recently studied \cite{Fernandez-Fontelo2016}.

There is an enormous global concern around 2019-novel coronavirus (SARS-CoV-2) infection in the last few months, leading the World Health Organization (WHO) to declare public health emergency \cite{Sohrabi2020}. As the symptoms of this infection can be easily confused with those of similar diseases like Middle East Respiratory Syndrome Coronavirus (MERS-CoV) or Severe Acute Respiratory Syndrome Coronavirus (SARS-CoV), its incidence has been notably underreported, especially at the beginning of the outbreak in Wuhan (Hubei province, China) by December 2019.

\section{Model definition}\label{model}
Consider an unobservable process with an AutoRegressive Moving Average ($ARMA(p, r)$) structure defined by
\begin{equation}\label{eq:ARMA}
  X_t = \alpha_1 X_{t-1} + \ldots + \alpha_p X_{t-p} + \theta_1 \epsilon_{t-1} + \ldots + \theta_r \epsilon_{t-r} + \epsilon_t,
\end{equation}
where $\epsilon_t$ is a Gaussian white noise process with $\epsilon_t \sim N(\mu_{\epsilon}, \sigma_{\epsilon}^2)$. The ARMA processes belong to the family of so called linear processes. Their importance relies on the fact that any stationary nondeterministic process can be written as a sum of a linear process and a deterministic component \cite{Brockwell1991}. These models are very well known, have been used in many applications since their introduction in the early 1950's and are general and flexible enough to be useful in a wide range of different contexts. Most used statistical software packages include functions that allow straightforward fitting of this family of models, so it seems a natural choice in the present work.

In our setting, this process $X_t$ cannot be directly observed, and all we can see is a part of it, expressed as
\begin{equation}\label{eq:model}
    Y_t=\left\{
                \begin{array}{ll}
                  X_t \text{ with probability } 1-\omega \\
                  q \cdot X_t \text{ with probability } \omega
                \end{array}
              \right.
\end{equation}
The interpretation of the parameters in Eq.~(\ref{eq:model}) is straightforward: $q$ is the overall intensity of misreporting (if $0 < q < 1$ the observed process $Y_t$ would be underreported while if $q > 1$ the observed process $Y_t$ would be overreported). The parameter $\omega$ can be interpreted as the overall frequency of misreporting (proportion of misreported observations).

\subsection{Model properties}
Consider that the unobserved process $X_t$ follows an $ARMA(p, r)$ model as defined in Eq.~(\ref{eq:ARMA}). As can be seen in \textit{Appendix 1} (Supplementary Material), the observed process has mean $\mathbb{E}(Y_t) = \frac{\mu_{\epsilon}}{1-\alpha_1 - \ldots - \alpha_p} \cdot \left(1 - \omega + q \cdot \omega \right)$ and variance $\mathbb{V}(Y_t) = \left(\left(\frac{\sigma_{\epsilon}^2 \cdot (1+\theta_1^2+ \ldots + \theta_r^2)}{1-\alpha_1^2- \ldots - \alpha_p^2}\right)+\frac{\mu^2_{\epsilon}}{(1-\alpha_1 - \ldots - \alpha_p)^2}\right) \cdot (1+\omega \cdot (q^2-1))
-\frac{\mu^2_{\epsilon}}{(1-\alpha_1 - \ldots - \alpha_p)^2} \cdot (1-\omega+q \cdot \omega)^2$. The autocorrelation function of the observed process can be written in terms of the features of the hidden process $X_t$ as
 \begin{equation}\begin{array}{ll}\label{corr_Y}
 	\rho_{Y}(k) & = \frac{V(X_t) (1-\omega+q \omega)^2}{(V(X_t)+E(X_t)^2) (1+\omega (q^2-1))-E(X_t)^2 (1-\omega + q \cdot \omega)^2} \cdot \rho_{X}(k) =\\
 & = c(\alpha_1, \ldots, \alpha_p, \theta_1, \ldots, \theta_r, \mu_{\epsilon}, \sigma^2_{\epsilon}, \omega, q) \cdot \rho_X(k),
\end{array}\end{equation}
where $\rho_X$ is the autocorrelation function of the unobserved process $X_t$.

A situation of particular interest is the case $\omega = 1$, meaning that all the observations might be 
underreported and that a simpler model for $Y_t$ excluding the parameter $\omega$ 
might be suitable
\begin{equation}
 Y_t = q \cdot X_t.
\end{equation}

In this case, however, the observed process $Y_t$ would be a non-identifiable $ARMA(p, r)$ model as the parameter $q$ cannot be estimated on the basis of the methodology described in the following section.

\subsection{Estimation\label{estimation}}
The likelihood function of the observed process $Y_t$ is not easily computable but the parameters of the model can be estimated by means of an iterative algorithm based on its marginal distribution, using the 
R packages \textit{mixtools} \cite{Benaglia2009a} and \textit{forecast} \cite{Hyndman2008, Hyndman2018}. The main steps are described in detail below:
\begin{enumerate}
 \item[(i)] Following Eq.~(\ref{eq:model}), the observed process $Y_t$ can be written as $Y_t = (1-Z_t) \cdot X_t + q \cdot Z_t \cdot X_t$, where $Z_t$ is an indicator of the underreported observations, following a Bernoulli distribution with probability of success $\omega$ $(Z_t \sim Bern(\omega))$. The marginal distribution of $Y_t$ is a mixture of two normal random variables  
 $N \left(\mu, \sigma^2 \right)$ and $N \left(q \cdot \mu, q^2 \cdot \sigma^2 \right)$ respectively, where $\mu = \frac{\mu_{\epsilon}}{1-\alpha_1-\ldots-\alpha_p}$ and $\sigma^2=\frac{\sigma_{\epsilon}^2 \cdot (1+\theta_1^2 + \ldots + \theta_r^2)}{1-\alpha_1^2- \ldots - \alpha_p^2}$. This fact can be used to obtain initial estimates for $q$ and $\omega$. Using the EM algorithm (specifically on the E-step), the posterior probabilities (conditional on the data and the obtained estimates) can be computed. This can be done using, for instance, the R package \textit{mixtools}.
\item[(ii)] Using the indicator $\hat{Z_t}$ obtained in the previous step, the series is divided in two: One including the underreported observations (treating the non-underreported values as missing data) and another with the non underreported observations (treating the underreported values as missing data). An $ARMA$ model is fitted to each of these two series and a new $\hat{q}$ is obtained by dividing the fitted means. 
 \item[(iii)] A mixture of two normals is fitted to the observed series $Y_t$ with mean and standard deviation fixed to the corresponding values obtained from the previous step, and a new $\omega$ is estimated.
 \item[(iv)] Steps (ii) and (iii) are repeated until the quadratic distance between two consecutive iterations $(\hat{q}_i-\hat{q}_{i-1})^2+(\hat{\omega}_i-\hat{\omega}_{i-1})^2+\sum_j (\hat{\alpha}_{j_{i}}- \hat{\alpha}_{j_{i-1}})^2 + \sum_k (\hat{\theta}_{k_{i}}- \hat{\theta}_{k_{i-1}})^2$ is below a fixed tolerance level.
 \item[(v)] Once the parameter estimates are stable according to the previous criterion, the underlying process $X_t$ is reconstructed as $\hat{X_t}=(1-\hat{Z_t}) \cdot Y_t + \frac{1}{\hat{q}} \cdot \hat{Z_t} \cdot Y_t$, and an $ARMA$ model is fitted to the reconstructed process to obtain $\hat{\alpha}_j$, $j=1, \ldots, p$, $\hat{\theta}_k$, $k=1, \ldots, r$ and $\hat{\sigma_{\epsilon}}^2$.
\end{enumerate}
To account for potential trends or seasonal behaviour, covariates can be included in the described estimation process. Additionally, a parametric bootstrap procedure with 500 replicates is used to estimate standard errors and build confidence intervals based on the percentiles of the distribution of the estimates. The R codes used to estimate the parameters and build the confidence intervals are available from the authors upon request.

\section{Results\label{results}}
The results of the proposed methodology over a comprehensive simulation study and an application on two real data sets are shown in this Section.

\subsection{Simulation study\label{simulation}}
A thorough simulation study has been conducted to ensure that the model behaves as expected, including $AR(p)$, $MA(r)$ and $ARMA(p, r)$ for $1 \leq p, r \leq 3$ structures for the hidden process $X_t$ with values for the parameters $\alpha$, $\theta$, $q$ and $\omega$ ranging from 0.1 to 0.9 for each parameter (some combinations of parameters have been omitted for $p>1$ or $r>1$ to ensure stationarity). For $ARMA(p, r)$ structures with $p>1$ or $r>1$ the parameters covered the same range (0.1 to 0.9) but with a difference of 0.2 instead of 0.1 for computational feasibility. Only average absolute bias, interval coverage and 95\% confidence interval corresponding to $p=r=1$ are shown in Table~\ref{tab:estim_sim}, as higher order models behave in a very similar manner (see Supplementary Material for details). These values are averaged over all combinations of parameters. Additionally, standard $AR(1)$, $MA(1)$ and $ARMA(1, 1)$ models were fitted to the same simulated series without accounting for their underreporting structure.

For each autocorrelation structure and parameters combination, a random sample of size $n = 1000$ has been generated using the function \textit{arima.sim} from R package \textit{forecast} \cite{Hyndman2008, Hyndman2018}. Different sample sizes ($n = 50, 100, 500$) have also been considered to study the impact of sample size on accuracy and the results are reported in the Supplementary Material. The performance of the proposed methodology is summarised in Tables S1 to S4 for $n = 50, 100, 500$ and $1000$ respectively. Average absolute bias is similar regardless of the sample size, while average interval lengths (AIL) are higher and interval coverages are poorer (around 75\% for $n = 50$) for lower sample sizes as could be expected. Several bootstrap sizes ($b = 20, 50, 100, 500$) were also considered and the difference between them were negligible, so only results corresponding to $b = 500$ bootstrap replicates are reported. 

\begin{table}[h!]
\centering
\caption{Model performance measures (average absolute bias, average interval length and average coverage) summary based on a simulation study.}
\label{tab:estim_sim}
\begin{tabular}{|p{2cm}|c|c|c|c|}
\hline
Structure & Parameter & Bias & AIL & Coverage (\%)\\
\hline
\multirow{3}{*}{$AR(1)$} & $\hat{\alpha}$ & 0.003      & 0.101      & 94.79\% \\
                         & $\hat{q}$      & $<10^{-3}$ & 0.001      & 93.28\% \\
                         & $\hat{\omega}$ & $<10^{-3}$ & 0.057      & 92.73\% \\
Standard $AR(1)$         & $\hat{\alpha}$ & 0.500      & 0.124      & 0.96\%  \\                
\hline
\multirow{3}{*}{$MA(1)$} & $\hat{\theta}$ & 0.001      & 0.116      & 95.61\% \\
                         & $\hat{q}$      & $<10^{-3}$ & $<10^{-3}$ & 92.87\% \\
                         & $\hat{\omega}$ & $<10^{-3}$ & 0.055      & 93.96\% \\
Standard $MA(1)$         & $\hat{\theta}$ & 0.499      & 0.124      &  1.23\% \\                
\hline
\multirow{4}{*}{$ARMA(1, 1)$} & $\hat{\alpha}$ & 0.003          & 0.170 & 95.50\% \\
                              & $\hat{\theta}$ & 0.006          & 0.213 & 96.77\% \\
                              & $\hat{q}$      & $<10^{-3}$     & 0.004 & 95.61\% \\
                              & $\hat{\omega}$ & $<10^{-3}$     & 0.065 & 94.35\% \\
\multirow{2}{1cm}{Standard $ARMA(1, 1)$}       & $\hat{\alpha}$ & 0.492 & 3.056 & 52.48\% \\ 
                                               & $\hat{\theta}$ & 0.509 & 3.055 & 51.14\% \\
\hline
\end{tabular}
\end{table}

It is clear from Table~\ref{tab:estim_sim} that ignoring the underreported nature of data (labeled as \textit{Standard} models in the table) leads to highly biased
estimates with extremely low coverage rates, even with larger average interval lengths. This is especially relevant
when the intensity or frequency of underreported observations is high.

\subsection{Example: HPV infection incidence\label{example}}
\begin{figure}
\centering
\includegraphics[height=0.46\textwidth]{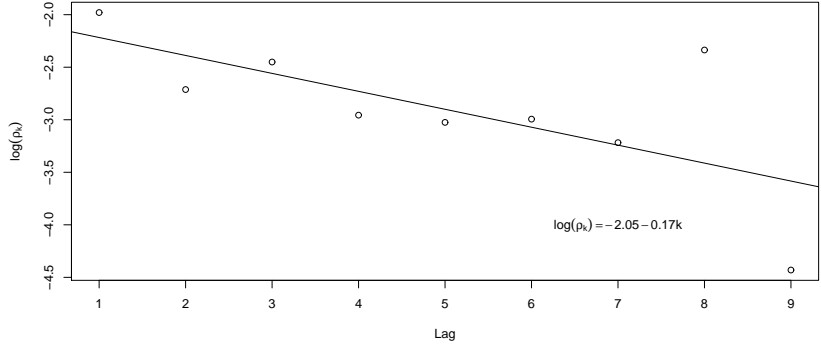}
\caption{Sample autocorrelation coefficients and estimated regression line of $\log(\rho_Y(k)) = \log(c)+k \cdot \log(\alpha)$.\label{fig:log_rho_k}}
\end{figure}

The series of weekly cases of HPV infection in Girona in the period 2010-2014 was previously 
analysed as a discrete $INAR(1)$ hidden Markov process \cite{Fernandez-Fontelo2016}. In a similar 
way, we aim to analyse the corresponding series of incidence, and an AR process of 
order 1 seems to be adequate (see Figure~\ref{fig:log_rho_k}). Additionally, the $AR(1)$ structure has the lowest AIC when compared to similar alternative models like $AR(2)$, $ARMA(1, 1)$ and $MA(1)$ (AICs are 291.63, 292.78, 292.78 and 292.02 respectively).
According to Eq.~(\ref{corr_Y}), the autocorrelation function of the observed process $Y_t$ when the hidden process $X_t$ has an $AR(1)$ structure takes the form $\rho_Y(k) = c \cdot \alpha^k$, where \\
$c = c(\alpha, \mu_{\epsilon}, \sigma^2_{\epsilon}, \omega, q) = \frac{(1-\omega+q \cdot \omega)^2 \cdot \sigma^2_{\epsilon}}{(1-\alpha^2) \cdot \left(\left( \frac{\sigma^2_{\epsilon}}{1-\alpha^2}+\frac{\mu^2_{\epsilon}}{(1-\alpha^2)}\right) \cdot (1+\omega \cdot (q^2-1)) - (1-\omega+q \cdot \omega)^2 \cdot \frac{\mu_{\epsilon}^2}{(1-\alpha)^2} \right)}$. In particular, in this case we can write $\log(\rho_Y(k)) = \log(c)+k \cdot \log(\alpha)$, so a statistically significant intercept of this linear regression model (estimating the parameters by ordinary least squares method) could be interpreted as an evidence of underreporting, as in this case ($p-value = 0.0016$). It is clear from Figure~\ref{fig:log_rho_k} that the estimated regression line does not cross the origin, so the behavior of the observed process is consistent with an underlying underreported $AR(1)$ process. 

By means of the estimation method described in Section \textit{Estimation}, it can be seen that 
the estimated model for the hidden process is $X_t = 0.109 \cdot X_{t-1} + \epsilon_t$, 
being the observed process $Y_t$,

\begin{equation} Y_t = \begin{cases}
      X_t & \text{with probability } 0.240 \\
      0.242 \cdot X_t & \text{with probability } 0.760
   \end{cases}
\end{equation}

The estimated parameters are reported in Table~\ref{tab:estimHPV}. 

\begin{table}[h!]
\centering
\caption{Bootstrap means and standard errors of the proposed model for the HPV example.}
\label{tab:estimHPV}
\begin{tabular}{|c|c|c|}
\hline
Parameter & Bootstrap mean & Bootstrap SE\\
\hline
$\hat{\mu_{\epsilon}}$ & 0.560 & 0.079\\
$\hat{\alpha}$         & 0.109 & 0.044\\
$\hat{\omega}$         & 0.760 & 0.154\\
$\hat{q}$              & 0.242 & 0.034\\
\hline
\end{tabular}
\end{table}

These results are highly consistent with those previously reported in the literature for the number of HPV
cases obtained through a discrete time series approach \cite{Fernandez-Fontelo2016} and can 
be interpreted in a very straightforward way. Moreover, this new methodology can be used to model the incidence of the disease instead of the number of cases, accounting for potential changes in the underlying population.

\begin{figure}
\centering
\includegraphics[height=0.5\textwidth, width=0.7\textwidth]{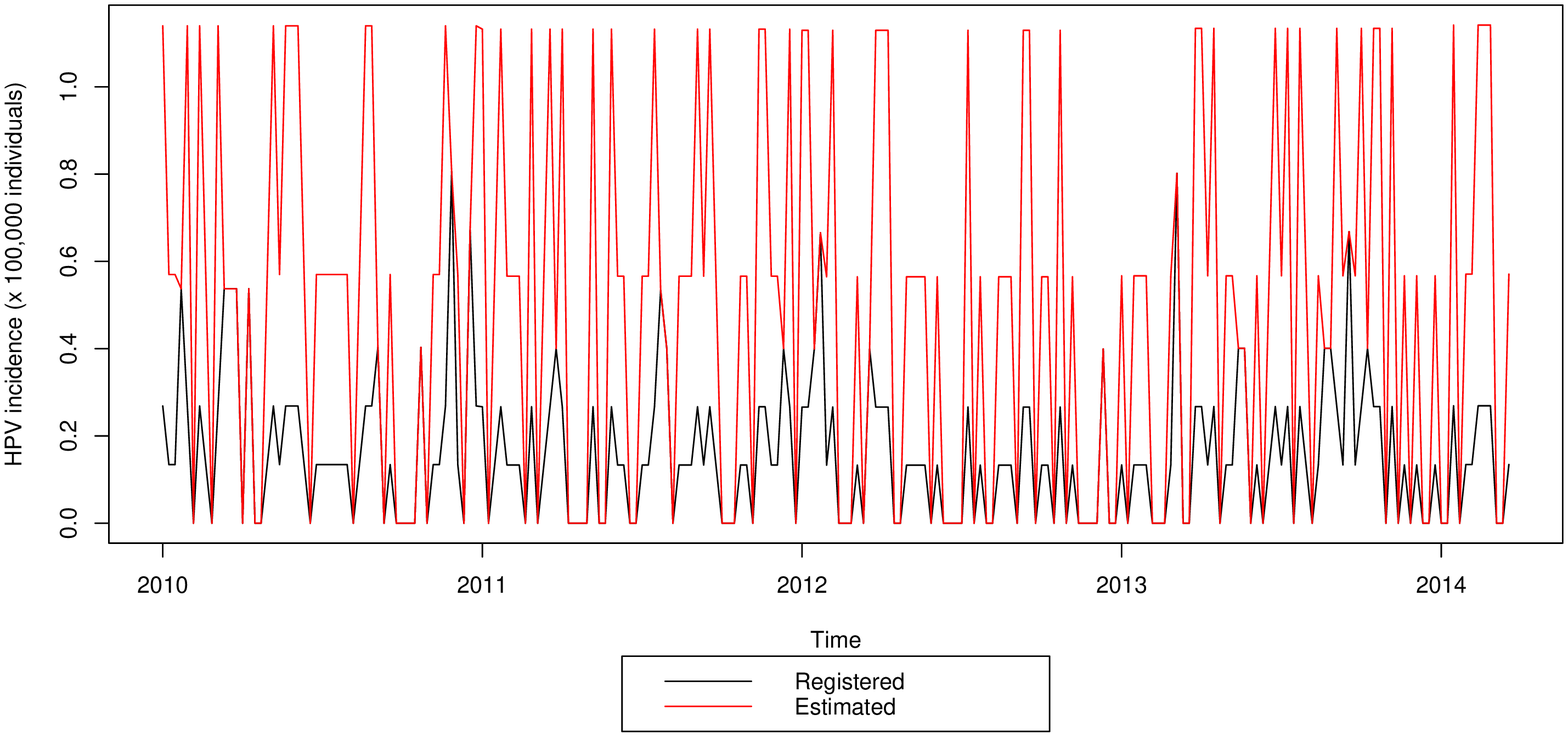}
\caption{Registered and estimated HPV incidence in Girona 
	in the period 2010-2014.
\label{fig:HPV_series}}
\end{figure}

The estimated intensity of underreporting is $\hat{q} = 0.242$, with 95\% confidence interval
$(0.176, 0.308)$. The registered and estimated evolution of HPV incidence within the 
study period (2010-2014) can be seen in Figure~\ref{fig:HPV_series}.

These results indicate that only 34\% of the HPV incidence in the considered period of time was actually recorded. Taking into account that public health cervical cancer prevention strategies are often designed on the basis of simulation models which are calibrated to registered HPV data \cite{Morina2017}, it is clear that providing decision makers with accurate data on HPV incidence is key to ensure optimal allocation of scarce public health funds.  

\subsection{Example: COVID-19 incidence in the region of Heilongjiang\label{example2}}
\begin{figure}
\centering
\includegraphics[height=0.45\textwidth, width=0.7\textwidth]{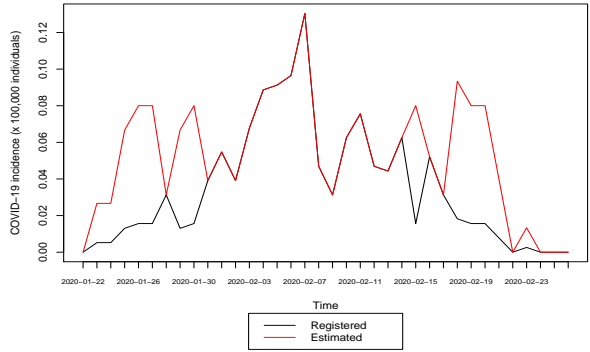}
\caption{Registered (black) and estimated (red) COVID-19 incidence in the region of Heilongjiang in the period 2020/01/22-2020/02/26.
\label{fig:COVID_series}}
\end{figure}

The betacoronavirus SARS-CoV-2 has been identified as the causative agent of an unprecedented world-wide outbreak of pneumonia starting in December 2019 in the city of Wuhan (China) \cite{Sohrabi2020}, named as COVID-19. Considering that many cases run without developing symptoms beyond those of MERS-CoV, SARS-CoV or pneumonia due to other causes, it is reasonable to assume that the incidence of this disease has been underregistered, especially at the beginning of the outbreak \cite{Zhao2020}. This work focuses on the COVID-19 incidence registered in Heilongjiang province (north-eastern China) in the period (2020/01/22-2020/02/26), and it can be seen in Figure~\ref{fig:COVID_series} that the registered data (black color) reflect only a fraction of the actual incidence (red color).

\begin{figure}
\centering
\includegraphics[height=0.45\textwidth]{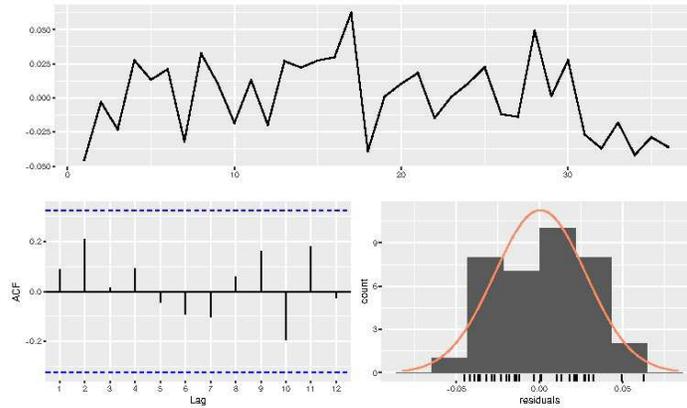}
\caption{Residual analysis (raw residuals, ACF and histogram) from a $MA(1)$ model.
\label{fig:COVID_residuals}}
\end{figure}

A disease with a similar behavior (MERS-CoV) has been modeled in a previous work as an $ARMA(3,1)$ \cite{Alkhamis2019}, so we evaluated the performance of this model and similar ones. Probably due to the shortness of the available data this autoregressive structure was not observed and in our case the best performing model was an $MA(1)$ (AIC of -151.04 against -148.49 for the $ARMA(3, 1)$), consistently with the residuals profile shown in Figure~\ref{fig:COVID_residuals}, obtained from fitting an $MA(1)$ model to the most likely process $X_t$ reconstructed following step (v) in Section \textit{Estimation}. 

By means of the estimation method described in Section \textit{Estimation}, it can be seen that the estimated model for the hidden process is $X_t = 0.481 \cdot \epsilon_{t-1} + \epsilon_t$, 
being the observed process $Y_t$,

\begin{equation} Y_t = \begin{cases}
      X_t & \text{with probability } 0.483 \\
      0.194 \cdot X_t & \text{with probability } 0.517
   \end{cases}
\end{equation}

The estimated parameters are reported in Table~\ref{tab:estimCOVID}. 

\begin{table}[h!]
\centering
\caption{Bootstrap means standard errors of the proposed model for the Covid-19 example.}
\label{tab:estimCOVID}
\begin{tabular}{|c|c|c|}
\hline
Parameter & Bootstrap mean & Bootstrap SE\\
\hline
$\hat{\mu_{\epsilon}}$    & 0.057 & 0.014\\
$\hat{\theta}$ & 0.481 & 0.196\\
$\hat{\omega}$ & 0.517 & 0.180\\
$\hat{q}$      & 0.194 & 0.094\\
\hline
\end{tabular}
\end{table}

\section{Discussion\label{discussion}}
In biomedical and epidemiological research, the usage of disease registries in order to analyse the impact and incidence of health issues is very common. However, the accuracy and data quality of such registries is in many cases at least doubtful. This is the case, for instance, for  rare diseases \cite{Kodra2017} or health issues that clear asymptomatically in most cases like HPV infection. In the case of HPV incidence in Girona in the period 2010-2014, the registered weekly average is 0.17 cases per 100,000 individuals, while the reconstructed process has a weekly average of 0.50 cases per 100,000 individuals, showing that only 34\% of the real incidence is recorded by the public health system. It must be considered that HPV infection is related to subsequent complications such as cervical cancer in some cases and that public health cervical cancer prevention strategies are often designed on the basis of simulation models which are calibrated to registered HPV data \cite{Morina2017} and therefore the optimal allocation of scarce public health resources cannot be ensured if the under-reporting issue is not accounted for. This result is very consistent with that of \cite{Fernandez-Fontelo2016}, where the authors claim that only 38\% of the HPV cases were registered in the same area and period of time. 

The Heilongjiang region COVID-19 data reveal that in average about 66\% of the actual incidence in the period 2020/01/22-2020/02/26 was reported. The unavailable data estimated by the proposed methodology are crucial to provide public health decision-makers with reliable information, which can also be used to improve the accuracy of dynamic models aimed to estimate the spread of the disease \cite{Zhao2020}. In China and almost globally afterwards, different non-pharmaceutical interventions were undertaken in order to minimise the impact of the disease on the general population and especially over the health systems, which were put to the limit of their capacity by the pandemic. In this context, one of the main challenges in predicting the evolution of the disease or evaluating the impact of these strategies is to use data as accurate as possible, taking into account that many COVID-19 cases are asymptomatic or with mild symptoms and a generalized shortage of testing kits \cite{Huang2020}, and therefore knowing that the registered number of affected individuals might be severely underestimated. 

The concerns around accuracy of registered data have recently led to the publication of recommendations to improve data collection to ensure accuracy of registries (see for instance \cite{Kodra2018,Harkener2019}). Nonetheless, these recommendations are very recent and may be difficult for the public health services of many countries to fully implement them, due to operational or structural issues.

The proposed methodology is able to deal with underreported (or overreported) data in a very natural and straightforward way, estimating its intensity and frequency on a continuous time series, and allowing to reconstruct the most likely unobserved process. It is also flexible enough to handle covariates  straightforwardly, and therefore it is simple to introduce trends or seasonality if necessary, so it can be useful in many contexts, where these issues might arise.  

The simulation study shows that the proposed methodology behaves as expected and that the parameters used in the simulations, under different autocorrelation structures, are properly recovered, regardless of the intensity and frequency of the underreporting issues.

\section*{Acknowledgments}
This work was supported by grant COV20/00115 from Instituto de Salud Carlos III (Spanish Ministry of Health). This work was partially supported by grant RTI2018-096072-B-I00 from the Spanish Ministry of Science and Innovation.

\bibliographystyle{plain}
\bibliography{refs}

\end{document}